\documentclass[12pt]{iopart}
\usepackage{graphicx}
\def\Journal#1#2#3#4{#4 {#1} {\bf #2} #3}

\def\PRB{Phys. Rev. B}

\def\Science{Science}
\def\JPL{J. Appl. Phys.}
\def\APL{Appl. Phys. Lett.}

\def\JPC{J. Phys. C}
\def\Nature{Nature}

\def\PhysicaE{Physica E}
\def\JACS{J. Am. Chem. Soc.}
\newcommand{\n}{\nonumber}
\newcommand{\bn}{\begin{eqnarray}}
\newcommand{\en}{\end{eqnarray}}
\newcommand{\eml}{\end{multline}}
\newcommand{\bml}{\begin{multline}}
\newcommand{\h}{\hspace}

\begin{document}

\title {Radial Dependence of the Carrier Mobility in Semiconductor Nanowires}
 \author{Kunal K. Das and Ari Mizel}
\address{Department of Physics, The Pennsylvania State
University, \\University Park, Pennsylvania 16802}

\ead{kdas@phys.psu.edu, ari@phys.psu.edu}

\begin{abstract}
The mobility of charge carriers in a semiconductor nanowire is
explored as a function of increasing radius, assuming low
temperatures where impurity scattering dominates.  The competition
between increased cross-section and the concurrent increase in
available scattering channels causes strongly non-monotonic
dependence of the mobility on the radius. The \emph{inter}-band
scattering causes sharp declines in the mobility at the wire radii
at which each new channel becomes available. At intermediate radii
with the number of channels unchanged the mobility is seen to
maintain an exponential growth even with multiple channels. We
also compare the effects of changing the radial scaling of the
impurity distribution.  We use transverse carrier wavefunctions
that are consistent with boundary conditions and demonstrate that
the $\delta$-function approximate transverse profile leads to
errors in the case of remote impurities.
\end{abstract}

\section{Introduction\label{intro}}

The relentless miniaturization of devices has made the physics of
lower dimensions a commonplace experimental reality today. A
variety of materials are routinely fabricated into nanometer scale
elements by research groups around the world;  at low
temperatures, nanoscale wires can physically behave as
one-dimensional conductors in the sense that the carriers are
confined to a single, or at most a limited number, of quantum
modes or channels in the transverse directions. The study of
transport in such one-dimensional conductors continues to reveal
novel physical behavior. In this context semiconducting nanowires
in particular have been the focus of sustained study because of
the ever-growing technological impact of semiconductors. Several
semiconducting materials are being employed to grow nanowires, and
to fabricate elementary devices, including silicon
\cite{Lieber-Si-1,Lieber-Si-2}, gallium arsenide
\cite{Samuelson-1,Hiruma}, germanium \cite{Yang, GeFET}, indium
phosphide \cite{Lieber-InP,InP-epitax} and indium arsenide
\cite{Samuelson-2}.

The carrier mobility is obviously a critical parameter
characterizing the transport properties and ultimately the
applications potential of a semiconductor nanowire. Experimental
studies frequently cite the observed mobility range for samples of
interest. For quasi one-dimensional wires made of a specific
semiconducting material, the mobility of the carriers depend on
several factors, such as the available scattering mechanisms for
the carriers, the carrier concentration, the temperature and the
physical dimensions of the wire.   Numerous theoretical papers
\cite{Sakaki,Spector,
Spector-2,Gold-Ghazali,Fishman,DasSarma,Weng-Leburton,Kirakosyan,Ari}
have studied the influence of essentially every relevant physical
parameter on the mobility in ultra small semiconductor wires.
However such studies have generally been in regimes where either
(i) the wires are thick enough and the temperatures sufficiently
high that the transport has bulk behavior, or (ii) they are in the
extreme size quantized limit with only one or two channels
available. Therefore it is of interest to see how the mobility in
a semiconductor nanowire behaves when the number of available
channels increases as the wire radius becomes larger and the
system deviates from the extreme size quantization limit.  That is
our goal in this paper. A strong practical motivation exists
because with current fabrication methods it is already possible to
grow nanowires with specific diameters with a precision of within
$10\%$ standard deviation \cite{Lieber-diam1,Korgel,Lieber-diam2},
and that is certain to improve rapidly with time.

In some earlier descriptions of mobility in wires at the size
quantized regime artificial transverse profiles were sometimes
used for the carrier concentration, chosen primarily for
mathematical convenience \cite{Spector,Fishman}. We treat the
carrier transverse motion quantitatively using carrier density
profiles that are consistent with physical boundary conditions and
considering multiple distinct transverse modes. Size quantization
is most pronounced in the regime of very thin wires and low
temperatures. Phonon scattering is suppressed at sufficiently low
temperatures, so in this paper we work in a regime in which the
mobility is mainly determined by impurity scattering. Some recent
studies \cite{Kirakosyan} have considered impurities concentrated
on the axis.  Here, we instead take the impurities to be
distributed outside the region of high carrier density since this
captures modulation doping \cite{Stormer} or surface roughening
effects resulting from some nanowire fabrication methods.

The rest of the paper is arranged as follows: In section
\ref{relax-times}, we develop the relaxation time description of
the mobility applied to one dimensional systems.  We specialize to
nanowires with a circular cross-section in section \ref{cylin} and
derive the impurity scattering matrix elements.  In section
\ref{surface}, we compute the radial scaling of those matrix
elements and of the carrier mobility assuming a surface
distribution of impurities. In section \ref{results}, we present
the results of our numerical calculations and our observations
regarding the behavior of the mobility and related properties as a
function of wire radius.

\section{Mobility in one dimension in terms of relaxation times\label{relax-times}}

The size quantization limit is most easily achieved at low
temperatures when most of the carriers have insufficient energy to
populate higher channels. The effect of phonon scattering is
negligible at sufficiently low temperatures and impurity
scattering is dominant. In order to focus exclusively on the
impurity scattering mechanism we consider the degenerate limit
corresponding to $T=0$. Our description will be based on the
relaxation time approximation to the Boltzmann equation that
approximates the collision term by the quotient of the deviation
from equilibrium of the Fermi distribution function and a
characteristic relaxation time. In quasi one dimension, the
distribution function $f_i({\bf k})$ carries a continuous label
$\bf k$ for the wavevector along the length of the wire and a set
of discrete \emph{channel} indices, denoted by $i$, that label the
degrees of freedom in the restricted transverse dimensions; the
same labelling is used for other physical parameters. Solving the
Boltzmann equation in the relaxation time approximation results in
a set of equations for the carrier relaxation times $\tau_i$ of
each channel \cite{Ferry-Goodnick} for a specific energy $E$
\bn \frac{1}{\tau_i(E)}=\sum_{j,{\bf k'}}S_{i,j}({\bf k},{\bf k'})
\left\{1-\frac{\tau_j(E)}{\tau_i(E)}\cos(\phi)\right\}.
\label{tau}\en
Here $S_{i,j}({\bf k},{\bf k'})$ is the scattering amplitude from
a state in channel $i$ with wavevector ${\bf k}$ to another in
channel $j$ with wavevector ${\bf k'}$.  The scattering mechanism
satisfies $S_{i,j}({\bf k},{\bf k'})=S_{j,i}({\bf k'},{\bf k})$.
The angle $\phi$ is between the initial $\bf k$ and the final $\bf
k'$ wavevectors.

\subsection{Scattering Matrix in One Dimension}

In one-dimension the wavevectors are scalars so that $ {\bf k}=\pm
k$ allowing only two possible values for the scattering angle,
$\phi=0,\pi$. The total energy of a carrier in the $i$th channel
has two parts
\bn E=\frac{\hbar^2}{2m}(k^2+\kappa_i^2), \h{1cm}
\epsilon_i(k)=\frac{\hbar^2}{2m}k^2. \label{parabolic}\en
The longitudinal energy $\epsilon_i(k)$ is assumed parabolic with the
lattice potential of the semiconductor incorporated into the
effective mass $m$ taken to be approximately the same for all
channels. The transverse energy $\hbar^2 \kappa_i^2/2m$ belongs to a
discrete spectrum of energies determined by the boundary conditions of
the transverse profile of the wire.

We carry out the sum over final momenta in Eq.~(\ref{tau}) using
the Fermi's Golden rule expression for the scattering amplitude
\bn \sum_{{\bf k'}}S_{i,j}({\bf k},{\bf k'})&=&\frac{L}{\hbar}\int
dk_j|\langle
k',j|V|k,i\rangle|^2\ \delta\left({\textstyle\frac{\hbar^2}{2m}(k^2+\kappa_i^2)-\frac{\hbar^2}{2m}(k'^2+\kappa_j^2)}\right)\n\\
&=&\sum_\pm\frac{mL}{\hbar^3k'}|\langle \pm
k',j|V|k,i\rangle|^2.\label{sumS}\en
Energy conservation determines the final wavevector given the
initial one and the specific channels involved, $k'
=\sqrt{k^2+(\kappa^2_i-\kappa_j^2)}$.

\subsection{Mobility in One Dimension}

A knowledge of the scattering matrix determines the relaxation
times which in turn determine the  electron mobility for each
channel in the wire through the expression
\bn \mu_i=\frac{q\sum_{k }v_k^2 [\tau_i(E)\partial_{\epsilon_i(k)}
f_i(k)]}{\sum_{k} f_i(k)}\en
where the carrier velocity for the parabolic case is $v_k=\hbar
k/m$. The derivative is with respect to the energy of the free
carriers.   With $E_F$ denoting the one-dimensional Fermi energy,
we can define Fermi energies for \emph{individual} channels
$E_i=E_F-\frac{\hbar^2}{2m}\kappa_i^2$ and the associated Fermi
wavevectors. Note that we leave out the label `$F$' for Fermi
surface quantities for individual channels as superfluous since,
as we see presently, we will only be working with Fermi surface
values for each channel.  At low temperatures where impurity
scattering is most pronounced, the Fermi distribution function is
essentially at the degenerate limit
\bn  f_i(k)=\theta(E_{i}-\epsilon_i(k)),\h{7mm}
\partial_{\epsilon_i(k)}f_i(k)=-\delta(E_{i}-\epsilon_i(k))\en
Allowing for spin degeneracy, the \emph{linear} density of
carriers in the $i$th channel is given by
\bn n_{Li}=2\int_{-\infty}^\infty
\frac{dk}{2\pi}\theta(E_{i}-\epsilon_i(k))
=\theta(E_{i})\frac{2k_{i}}{\pi} \label{nLi}\en
For a specific carrier density, $n=N/V$, and cross-section $A$
(assumed uniform) of the wire the linear density is $n_L=nA$. Then
the one-dimensional Fermi energy $E_F$ can be determined by adding
together the densities (\ref{nLi}) of each channel
\bn\label{lin-density} \frac{\pi}{2}n_L=\sum_i
\theta(E_F-{\textstyle
\frac{\hbar^2}{2m}}\kappa_i^2)\sqrt{{\textstyle
\frac{2m}{\hbar^2}}E_F-\kappa_i^2}\en
where $\theta(x)$ is the Heaviside unit step function. The
mobility in each channel becomes
\bn \mu_i&=&
-\theta(E_{i})\frac{2q\pi}{2k_{i}}\int_{-\infty}^\infty
\frac{dk}{2\pi}\frac{\hbar^2 k^2\tau_i(E)}{m^2}
\delta\left(E_{i}-\frac{\hbar^2k^2}{2m}\right)\n\\
&=&-\theta(E_{i})\frac{q\tau_i(E_F)}{m}\en
The average mobility for electrons is then given by
\bn \mu=\frac{\sum_i n_{Li}\mu_i}{\sum_i n_{Li}}=
\theta(E_{i})\frac{e}{m}\frac{\sum_i k_{i}\tau_i(E_F)}{\sum_i
k_{i}}\en
In the degenerate limit the relaxation times have to be evaluated
\emph{only} at the Fermi surface determined by the total carrier
density in the wire and the wire radius.

\section{Cylindrical nanowires\label{cylin}}

We will now specifically consider nanowires with an uniform
cylindrical cross-section, and a Coulomb scattering potential
arising from ionized impurities. At low temperatures where
impurity scattering is dominant a Coulomb potential is an
appropriate choice for the scattering potential $V$ in typical low
dimensional systems.  The natural basis functions are a product of
transverse functions involving Bessel functions and a plane wave
corresponding to the longitudinal part. So the matrix elements due
to an impurity at ${\bf r_0}\equiv \{z_0,r_0,\theta_0\}$ is
\bn \langle k_j,j|V({\bf r},{\bf r_0})|k_i,i\rangle &=&
\frac{e^2}{4\pi\epsilon L}\int_0^R\h{-2mm}rdr\!\int_0^{2\pi}
\h{-3mm}d\theta\
\Phi_j^*(r,\theta)\Phi_i(r,\theta)\n\\&&\times\int_{-\infty}^{\infty}
dz\frac{
e^{-i(k_j-k_i)z}}{\sqrt{(r^2+r_0^2-2rr_0\cos(\theta))+(z-z_0)^2}}\h{3mm}
\en
where the azimuthal angle $\theta$ is measured from the direction
of the impurity at $\theta_0$. Because of the cylindrical symmetry
the value of $\theta_0$ will not influence the scattering
probability. The quantity $\epsilon$ is the dielectric constant.
It is well known \cite{Spector-2} that due to screening of the
Coulomb interaction, in the strict degenerate limit of zero
temperature the static dielectric function in one dimension
evaluated in a random phase approximation (RPA) has a divergence
at twice the Fermi vector, $2k_i$, for a channel. From our
analysis above it is clear that for intra-channel scattering at
the degenerate limit the momentum change corresponds precisely to
that value, and therefore the dielectric constant also needs to be
evaluated at that divergent point. However our interest is in the
radial scaling of the mobility and not on evaluating its precise
value, so we adopt the assumption of low, but \emph{non-zero},
temperature used in a similar context in Ref. \cite{Fishman},
whereby the divergence is removed leading to a well-defined
dielectric function.  The analysis of Ref. \cite{Fishman} also
showed that a doubling of the wire radius caused relatively small
changes in the dielectric constant over a wide range of carrier
densities. But as we will establish in this paper, the effects of
increasing radius on the mobility on the other hand is exponential
in nature. Therefore the dependence of the dielectric function on
wire size should have little qualitative impact on the radial
scaling of the mobility. Hence we will treat the dielectric
function as a constant $\epsilon$ over the range of radii that we
consider in this paper.

The transverse wave functions involve Bessel functions and for
a wire of radius $R$ are given by
\bn
\Phi_i(r,\theta)=\frac{e^{il\theta}}{\sqrt{2\pi}}\frac{J_l(\kappa_{ln}
r)}{\frac{R}{\sqrt{2}}J_{l}'(\kappa_{ln}R)}.\en
The various channels are labelled by two indices $i\equiv{l,n}$,
with $l$ corresponding to the order of the Bessel functions
$J_l(\kappa_{ln}r)$, and $n$ labelling the zeroes for each order
in a sequence of increasing magnitude. The transverse eigenmodes
are determined by the boundary condition that the Bessel functions
vanish on the surface of the wire $J_l(\kappa_{ln}R)=0$, with the
zeroes denoted by $p_{ln}=\kappa_{ln}R$.

\begin{table}[t]
\caption{The zeroes $p_{ln}$ of the Bessel functions, for the
lowest transverse channels in a uniform cylindrical wire, are
listed along with the minimal values of the scaled wire radius
$n^{1/3}R$ at which each channel becomes populated with charge
carriers. For a bulk carrier density of $n=10^{24}$ m$^{-3}$ that
we use for numerical estimates, the right column would correspond
to radii in units of $10^{-8}\ {\rm m}=10$ nanometers. }
\begin{indented}\item[]
\begin{tabular}{rr}
\br
 $p_{\ {ln}}$ \h{6mm}\vline& $n^{1/3}R$ \\\mr
 $p_{0,1}=2.40482$\h{6mm} \vline& $0$ \\
 $p_{1,1}=3.83171$\h{6mm} \vline& $0.8455$ \\
 $p_{2,1}=5.13562$\h{6mm} \vline& $0.1173$ \\
 $p_{0,2}=5.52007$\h{6mm} \vline& $0.1305$ \\
 $p_{3,1}=6.38016$\h{6mm} \vline& $0.1539$ \\
 $p_{1,2}=7.01559$\h{6mm} \vline& $0.1706$ \\
 $p_{4,1}=7.58834$\h{6mm} \vline& $0.1856$ \\
 $p_{2,2}=8.41724$\h{6mm} \vline& $0.2048$\\
 $p_{0,3}=8.65372$\h{6mm} \vline& $0.2116$\\
 \br
\end{tabular}
\end{indented}
\end{table}

We assume a azimuthally symmetric layer of impurities of bulk density
$\rho(r_0)$ distributed between radii $r=a$ and $r=b$. On integrating over the impurity distribution we obtain
the scattering amplitude sum (\ref{sumS})

\bn\label{cyl-matrix-elem}\sum_{{\bf k'}}S_{j,i}({\bf k}',{\bf
k})&=&L\int_{a}^{b} dr_0 2\pi
r_0\rho(r_0)\sum_\pm\frac{mL}{\hbar^3k'}|\langle \pm
k_{l'n'},l'n'|V({\bf r},{\bf r_0})|k_{ln},ln\rangle|^2\n\\ &=&
S_{l'n',ln}^++S_{l'n',ln}^-\en
with
\bn \fl S_{l'n',ln}^\pm =\frac{me^4}{2\pi\epsilon^2 \hbar^3}\times
\frac{1}{k_{l'n'}}\int_{a}^{b} dr_0 r_0\rho(r_0)
\n\\\fl\h{5mm}\times\left|\int_0^R rdr\int_0^{2\pi}
\frac{d\theta}{2\pi} \frac{J_{l'}(\kappa_{l'n'} r
)J_l(\kappa_{ln}r )e^{i(l'-l)\theta}}{
\frac{R^2}{2}J_{l'}'(\kappa_{l'n'}R)J_{l}'(\kappa_{ln}R)}K_0\left(|(k_{ln}\mp
k_{l'n'}|\sqrt{r^2+r_0^2-2rr_0\cos(\theta)}\right)\right|^2. \n\en
The $+$ sign corresponds to forward scattering ($\phi=0$) and the
$-$ to backscattering ($\phi=\pi$). Using this notation in
Eq.~(\ref{tau}) gives a system of linear equations for the
relaxation times
 \bn\label{coupled-eqns}\fl \left[2S_{ln,ln}^{-}
+\sum_{\{l'n'\}\neq\{ln\}}\left(S_{l'n',ln}^{+} +
S_{l'n',ln}^{-}\right)\right]\tau_{ln}+\sum_{\{l'n'\}\neq
\{ln\}}\left(S_{l'n',ln}^{-}-S_{l'n',ln}^{+}\right)\tau_{l'n'} =
1.\en
At zero temperature the matrix elements, like the relaxation
times, are evaluated at the effective Fermi wavevector $k_{ln}$
for each channel. In the strict one dimensional limit when only
the lowest channel is available we retrieve the well known result
$\tau^{-1}=2S_{01,01}^{-}$ \cite{Spector}. It has been a common
practice to assume that in thin wires the carriers may be assumed
to be confined to the wire axis, thereby justifying the usage of
a $\delta$-function to approximate the transverse profile of the
carrier density, in which case the matrix elements in
Eq.~(\ref{cyl-matrix-elem}) reduce to
 \bn\label{delta-app} \h{-1mm}S_{l'n',ln}^\pm
\simeq\frac{me^4}{2\pi\epsilon^2 \hbar^3k_{l'n'}}
\h{-1mm}\int_{a}^{b} \h{-2mm}dr_0 r_0\rho(r_0)K_0^2[|k_{ln}\mp
k_{l'n'}|r_0]\en
While this is mathematically simpler, we will presently show that
this approximation is invalid for the cases we consider.

\section{Surface Impurity\label{surface}}

For some quasi-one dimensional nanowires, it is a good
approximation to treat the impurities as distributed in a layer of
varying thickness outside the wire; this models surface roughness
or modulation doping of dopants. We therefore take the scattering
centers to be distributed in a thin uniform layer of width $w$
along the surface of the wire. First we take the bulk density of
the impurities to be constant within that layer
$\rho(r_0)=\rho_B$. We then write the matrix elements in
Eq.(\ref{cyl-matrix-elem}) in a way that makes the radial
dependence more transparent:
\bn\label{common-factor}\fl S_{l'n',ln}^\pm =Q\times
\frac{(R/l_0)^2}{q_{l'n'}}\\\fl\h{1.5cm}\times\left|\int_0^1 xdx
\frac{J_{l'}(p_{l'n'} x )J_l(p_{ln}x)}{
\frac{1}{2}J_{l'}'(p_{l'n'})J_{l}'(p_{ln})}
\int_0^{2\pi}\frac{d\theta}{2\pi}e^{i(l'-l)\theta}K_0\left(|q_{l'n'}\mp
q_{ln}|\sqrt{1+x^2-2x\cos{\theta}}\right)\right|^2.\n\en
Everything that does \emph{not} depend on the radius has been
included in the pre-factor $Q$ which contains all the dimensioned
quantities. The rest of the expression contains only
\emph{dimensionless} quantities as we have rescaled the lengths by
the radius $R$, so that the integration variable is $x=r/R$ and
the wavevectors are $q=Rk$ . The scale and the dimension of the
matrix element are then set by the constant in front
\bn\label{Q} Q&=&\frac{me^4l_0^3n}{2\pi\epsilon_0^2
\hbar^3}\times\left[\frac{1}{\epsilon_s^2}\frac{m}{m_e}
\frac{\rho_B}{n}\frac{w}{l_0}\right]= 1.04\times10^{15}\ {\rm
s}^{-1}\times \left[\frac{1}{\epsilon_s^2}\frac{m}{m_e}
\frac{\rho_B}{n}\frac{w}{l_0}\right] \en
We have assumed a length scale of a nanometer $l_0=10^{-9}{\rm m}$
along the radial direction, for the wire radius $R$ as well as for
the width of the scattering layer $w$. We have also scaled the
impurity density $\rho_B$ by the carrier density $n$. The
dielectric constant of the wire is denoted by $\epsilon_s$ .

Another advantage of writing the matrix elements this way lies in
the fact that the factor $Q$ is common to all the matrix elements
$S_{l'n',ln}^{\pm}$ in the set of linear equations in
Eq.~(\ref{coupled-eqns}).  We can therefore divide through by that
factor, and since the multiplication of a column of a determinant
by a constant has the effect of multiplying the determinant by the
same constant, all the relaxation times $\tau_{ln}$ for every
channel carry a common factor of $Q^{-1}$.

In certain cases it is more accurate to assume that the linear density
of the impurities is constant instead of the bulk density of
impurities. In that case the radial scaling is somewhat different; we
then have to replace $w\rho_B\rightarrow \rho_L/(2\pi R)$ so that the
scattering matrix elements are

\bn\label{constant-linear} \fl S_{l'n',ln}^{'\pm}
=Q' \times \frac{(R/l_0)}{q'}\\\fl\h{1.5cm}\times\left|\int_0^1
xdx \frac{J_{l'}(p_{l'n'} x )J_l(p_{ln}x)}{
\frac{1}{2}J_{l'}'(p_{l'n'})J_{l}'(p_{ln})}
\int_0^{2\pi}\frac{d\theta}{2\pi}e^{i(l'-l)\theta}K_0\left(|q_{l'n'}\mp
q_{ln}|\sqrt{1+x^2-2x\cos{\theta}}\right)\right|^2.\n \en
The coefficient $Q'$ has the same scale factor as $Q$, but now the
bulk impurity density and the width of the impurity layer are
replaced by a linear density of the impurities $\rho_L$, scaled by
the bulk carrier density and a nanoscale area element $n\times
2\pi l_0^2$
\bn\label{Q-prime} Q'&=&\frac{me^4l_0^3n}{2\pi\epsilon_0^2
\hbar^3}\times\left[\frac{1}{\epsilon_s^2}\frac{m}{m_e}
\frac{\rho_L}{2\pi l_0^2n}\right].\en

\section{Results and Discussion\label{results}}

We now proceed with numerical estimates for gallium arsenide
(GaAs) for which our assumption of parabolicity (\ref{parabolic})
is appropriate. Since we are interested mainly in the radial
scaling we present our results scaled by the common constant $Q$
since it carries no radial dependency. But it determines the
intrinsic magnitude of the physical quantities, therefore we first
provide an estimate of its value for a GaAs nanowire. The
effective mass for GaAs is $m=0.068 m_e$, assumed the same for all
the channels, and the dielectric constant is about
$\epsilon_S=12$, not significantly altered by screening at the
high carrier density that we consider \cite{Fishman}. If we take
the impurity density to be equal to the carrier density and the
impurity layer to be of the order of a nanometer thick $w\simeq
10^{-9}$ m, we obtain a value of $ Q\simeq 4.9\times 10^{11}\
s^{-1}$. We may then estimate the magnitude of the average
mobility by writing it as
\bn \mu&=&\frac{e}{Qm}\theta(E_{i})\frac{\sum_i
q_{i}[Q\tau_i(q_{i})]}{\sum_i q_{i}}\en
The pre-factor contains all the dimensioned quantities and, for
the above-mentioned value of $Q$, it is $e/(Qm)\sim 5\times 10^4\
{\rm cm^2/V/s}$. Our numerical computation of the remaining
dimensionless part, presented below, then yields a mobility in the
range $10^4 - 10^7\ {\rm cm^2/V/s}$, which is consistent with the
magnitudes in experimental measurements.

\begin{figure}
\begin{center}\fl\includegraphics*[width=0.8\columnwidth,angle=0]{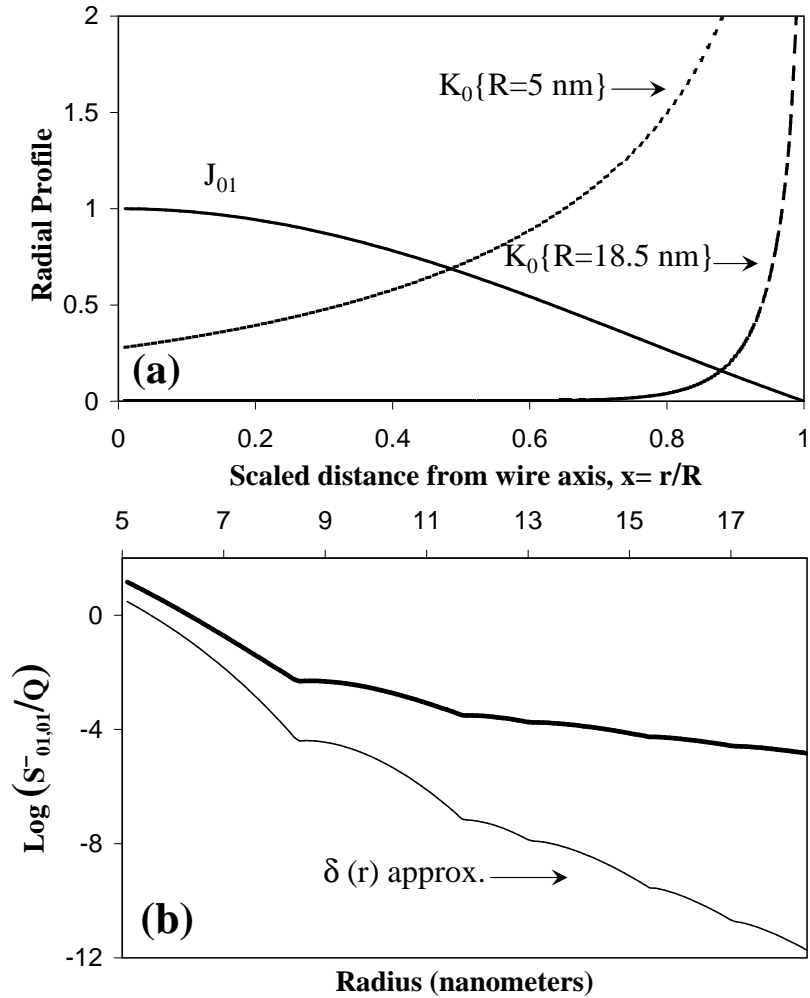}\vspace{-1.7cm}
\fl\caption{\label{fig1} (a) Plot of the radial profile of the
carrier distribution of the lowest channel, $J_{01}(p_{01}x)$,
shown alongside with the radial dependence of the impurity
scattering contained in the factors $K_0(2q_{01}|1-x|)$ shown for
the extreme values of the range of wire radii $R$ that we
consider. (b) Semi-log plot of the scattering matrix element,
scaled by the constant $Q$, for the lowest channel with (thin
line) and without (thick line) the $\delta$-function approximation
for the radial density of the carriers. }\end{center}
\end{figure}

\begin{figure}\begin{center}
\fl\includegraphics*[width=\columnwidth,angle=0]{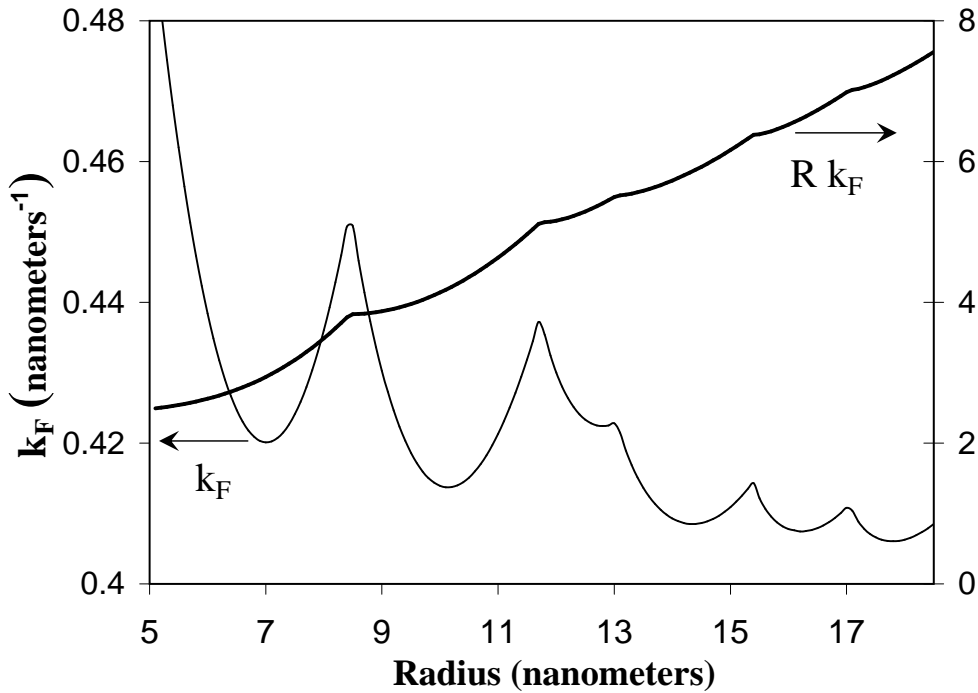}\vspace{-9.5cm}
\fl\caption{\label{fig2} The net Fermi wavevector
$k_F=\sqrt{2mE_F/\hbar^2}$, for the carriers in the wire for a
fixed carrier density $n=10^{24}\ {\rm m}^{-3}$, is plotted along
the left axis as a function of the radius. The dimensionless
quantity $Rk_F$ is plotted along the right axis.}\end{center}
\end{figure}

\begin{figure}\begin{center}
\fl\includegraphics*[width=\columnwidth,angle=0]{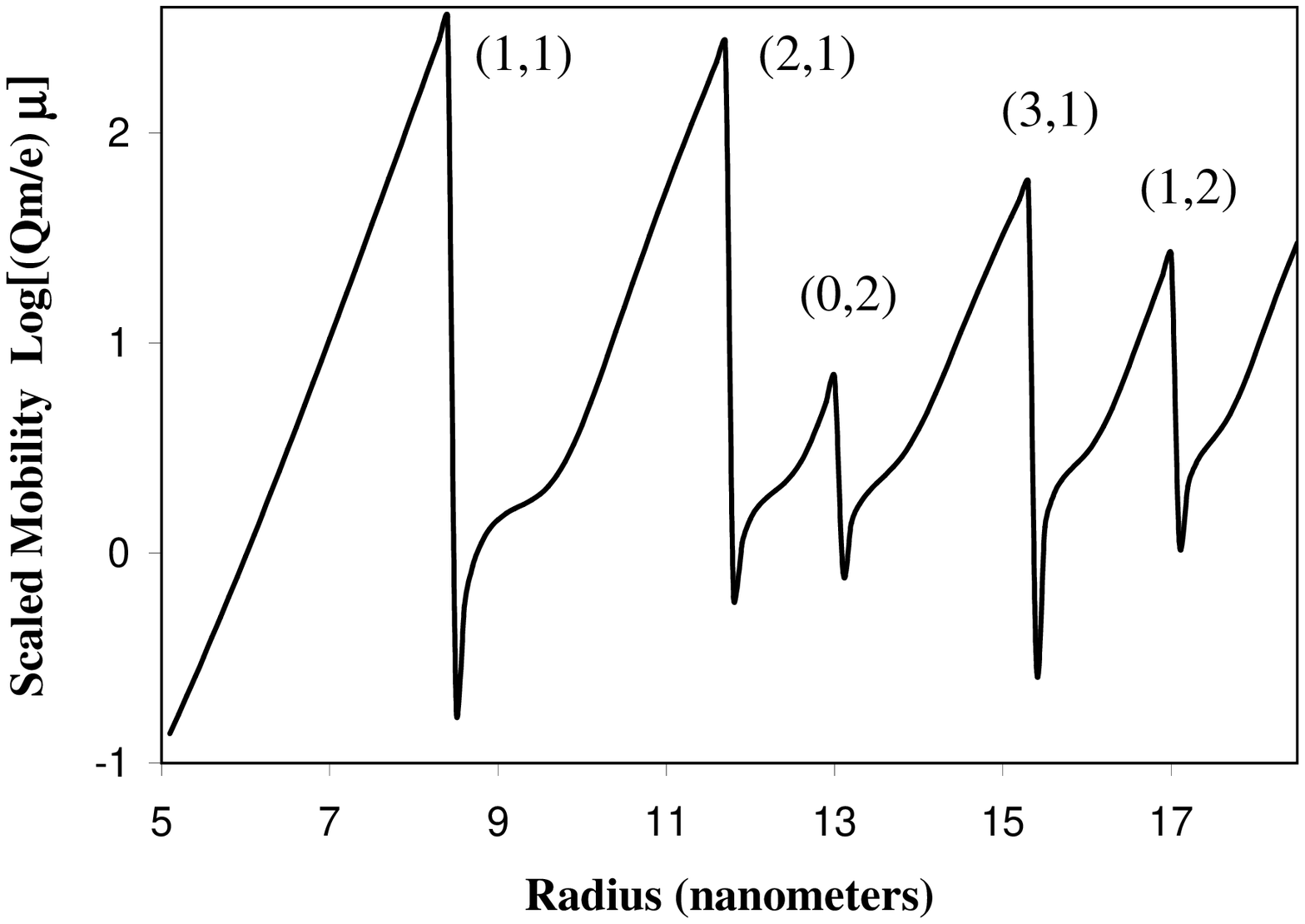}\vspace{-9.5cm}
\fl\caption{\label{fig3} The mobility scaled by $e/(Qm)$ is
plotted on a logarithmic scale as a function of the wire radius
for fixed carrier $n=10^{24}\ {\rm m}^{-3}$.  The $(l,n)$ label
for each new channel is shown as it becomes available causing the
sharp declines in the mobility. } \end{center}
\end{figure}

Having established the range of magnitudes of the mobility we now
turn our attention to the radial scaling. The numerical estimates
assume a carrier density of $10^{18}$ cm$^{-3}$ $=10^{24}$
m$^{-3}$. When all else is fixed the radius determines the number
of channels available for transport in Eq.~(\ref{lin-density}).
The minimal radii for each channel to be populated with carriers
having non-vanishing longitudinal energy are shown in Table~I.

In Fig.~\ref{fig1}(a) we plot the modified Bessel function of the
second kind for the lowest channel,
$K_0(x,\theta;q_{01})=K_0(2q_{01}\sqrt{1+x^2-2x\cos(\theta)})$; it
contains the effects of the Coulomb scattering in the scattering
matrix elements. In the $x$-$\theta$ plane, the function
$K_0(x,\theta;q_{01})$ is very strongly peaked along the $\theta=0$
line, therefore we specifically chose to plot $K_0(x,0;q_{01})$.
In order to see the behavior over the entire range of wire radii
we present plots for the minimum radius $R=5\ \rm{nm}$ and maximum
radius $R=18.5\ \rm{nm}$ that we consider. Alongside we plot the
un-normalized radial profile of the carrier density in the lowest
channel $J_{01}(p_{01}x)$.  We find that the \emph{overlap} of the
carrier density profile $J_{01}(p_{01}x)$ with the prominent
region of $K_0(x,\theta;q_{01})$ decreases significantly with
increased wire radius, indicating that an approximation that
replaces the radial profile with a delta function at the wire axis
would deteriorate rapidly with larger wire radius. That is exactly
what we see in Fig.~\ref{fig1}(b) where we plot the intra-band
scattering matrix element $S^-_{01,01}$ for the lowest channel
using first Eq.~(\ref{cyl-matrix-elem}) which uses the appropriate
radial profile and secondly Eq.~(\ref{delta-app}) which uses the
$\delta$-function approximation. The two curves are noticeably
different even for a wire radius of $5$ nm, but for larger wire
radii they differ by several orders of magnitudes. This is exactly
what one would expect for a surface distribution of impurities
because the impurities are further removed from the wire axis for a
larger wire.

Figure 2 shows the scaling of the Fermi wavevector with the
radius. We see that the dimensionless product of the wire radius
with the Fermi wavevector, $R\times k_F$ has an almost monotonic
growth with the radius. The Fermi vector itself decreases
noticeably as each new channel becomes available, and then
rebounds gradually but with an overall decline of the peak values
reached before each succeeding channel enters. So the general
trend is that $k_F$ gets smaller with increasing radius; with
sufficiently large number of channels we expect it to approach the
bulk value which, for the carrier density we have assumed, would
be $k_{F}^{3D}= (3\pi^2 10^{-24})^{1/3}\sim 0.31\ {\rm nm}^{-1}$.
Our plot suggests a gradual approach to that limit. That limit
gives a criterion for when the system makes the transition from
quasi-1D to 3D.

We plot the mobility in Fig. \ref{fig3} on a logarithmic scale to
show that the growth of the mobility with radius is of an
exponential nature in between points of sharp declines. The most
striking feature is that the multi channel scattering destroys the
simple monotonic growth of the mobility seen with a single channel
\cite{Spector}. As the radius increases, and each new channel
becomes energetically available for scattering there is a sharp
reduction in the average mobility. In between the addition of new
scattering channels, the increasing wire radius causes the
mobility to grow, but that growth has an inflection point,
reflecting the competition with increased scattering. The sharp
declines cause an overall lowering of the mobility as the radius
increases significantly. This is consistent with an eventual
approach to bulk behavior.

\begin{figure}\begin{center}
\fl\includegraphics*[width=\columnwidth,angle=0]{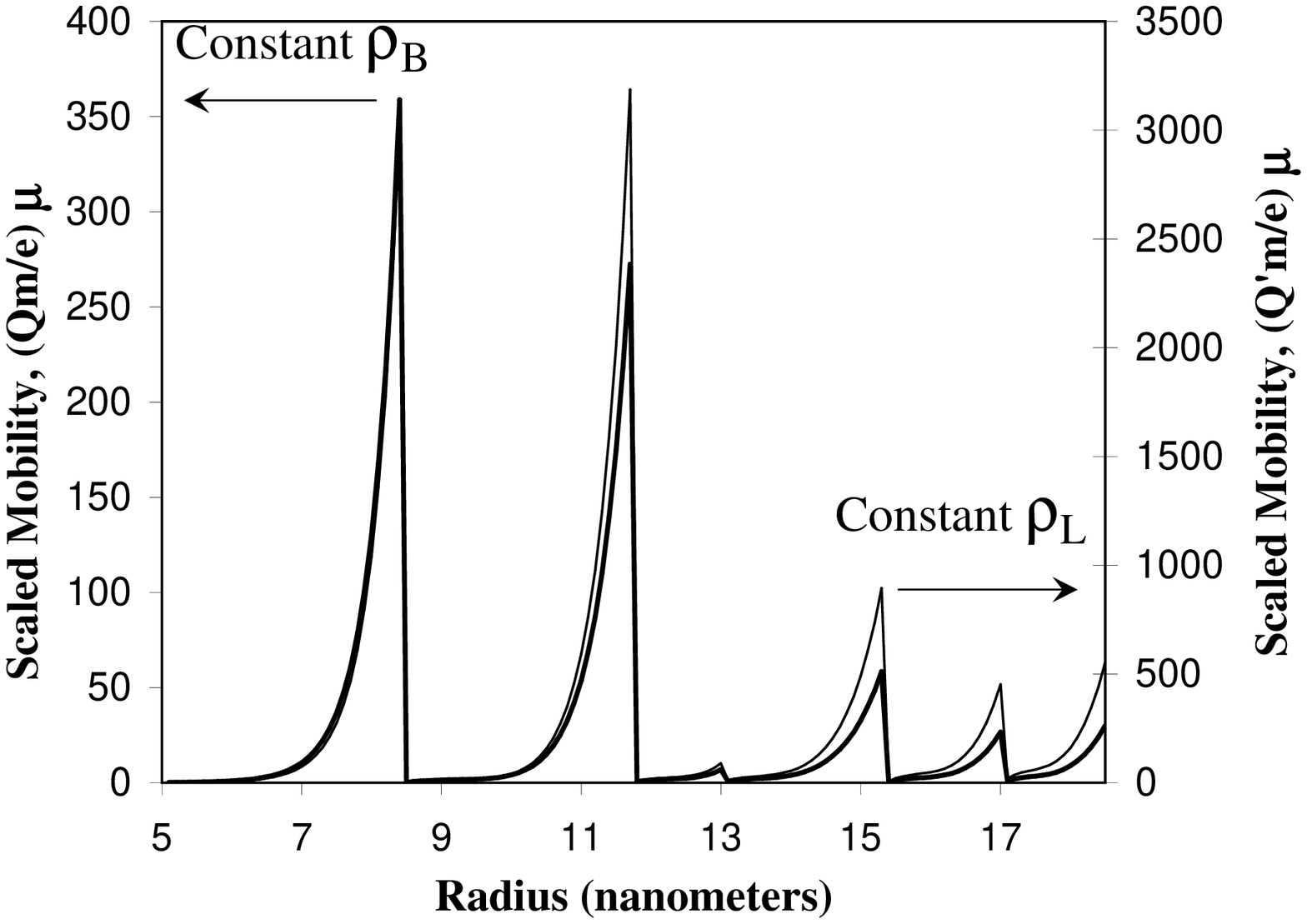}\vspace{-9.5cm}
\fl\caption{\label{fig4} Plot of the scaled mobility versus the
wire radius on a linear scale for constant bulk density  of
surface impurities (thick line, along the left axis) and for
constant linear density (thin line, along the right axis). Note
the scale factors are different for each, $Q$ and $Q'$ given by
Eqs.~(\ref{Q}) and (\ref{Q-prime}) respectively.} \end{center}
\end{figure}

We get a sense of the actual intrinsic magnitude of the mobility
and its variations in Fig.~\ref{fig4} where a linear scale is
employed. This figure also illustrates the effect of changing the
radial scaling of the impurity density itself;  we have plotted
alongside the mobility for the case described in
Eqs.~(\ref{constant-linear}) and (\ref{Q-prime}) where regardless
of the increased wire growth the linear density of the impurity
layer remains constant. Those equations show that there is then an
extra factor of $R$ in the mobility, because the surface density
becomes sparser with increased wire size; this causes the mobility
to increase faster at larger radii as we see from the plot.
Otherwise the two curves are quite similar implying that the
essential features are not affected significantly by the nature of
the impurity distribution, because the strong exponential behavior
of the modified Bessel functions $K_0$ dominates the trend.

We have assumed a simple model which highlights general trends in
the carrier mobility and related parameters as a semiconductor
nanowire deviates from the strict size quantization limit towards
bulk behavior. Real semiconductor wires will not have a uniform
cross section, and are likely to have impurities in the wire
interior as well as on the surface, so the behavior we have seen
will therefore likely be less sharp. Moreover it is known that
screening affects intra-band scattering more significantly
\cite{Weng-Leburton, Ferry-Goodnick} than inter-band scattering.
So a more accurate treatment of the dielectric function is likely
to diminish the changes in the mobility as new channels become
available. But the general trends we have established are
sufficiently striking that those features would be relevant in
experimentally observed behavior of the mobility as the wire size
is increased. In particular the effect of \emph{inter}-channel
scattering should be manifest in the degenerate limit.

\ack

The authors thank T. Mayer, S. Mohney, and J. Redwing for helpful
conversations.  We acknowledge the support of NSF NIRT program grant
DMR-0103068.

\end{document}